\documentclass[12pt]{article}
\usepackage{graphicx,amsmath,amssymb}
\usepackage{cite}
\usepackage{amsfonts}
\usepackage{amssymb}
\usepackage{amsmath}
\usepackage{multirow}
 
\usepackage{graphicx}  

\newcommand{\cL}{{\cal L}}

\newcommand{\nn}{\nonumber}
\newcommand{\crn}{\nn \\}
\newcommand{\degK}{^\circ\mathrm{K}~}
\newcommand{\power}[1]{\times 10^{#1}}

\newcommand{\Reg}{\mbox{Reg}}
\newcommand{\MPl}{M_\mathrm{Planck}}

\newcommand{\LPl}{\Lambda_{\rm Pl}}
\newcommand{\lear}{\leftarrow}
\renewcommand{\MPl}{M_{\rm Pl}}
\newcommand{\mpl}{M_{\rm Pl}}
\newcommand{\tpl}{t_{\rm Pl}}
\newcommand{\gv}{\mbox{GeV}}

\newcommand{\D}{\mathrm{d}}
\newcommand{\E}{\mathrm{e}}
\newcommand{\I}{\mathrm{i}}
\newcommand{\PT}{phase transition }

\newcommand{\veps}{\varepsilon}
\newcommand{\mbo}[1]{$#1$}
\newcommand{\semis}{\;;\;\;}

\newcommand{\comas}{\;,\;\;}

\newcommand{\MSb}{$\overline{\mathrm{MS}}$ }
\newcommand{\ba}{\begin{eqnarray*}}
\newcommand{\ea}{\end{eqnarray*}}
\newcommand{\bea}{\begin{eqnarray}}
\newcommand{\eea}{\end{eqnarray}}
\newcommand{\bary}{\begin{array}}
\newcommand{\eary}{\end{array}}

\newcommand{\sign}{\mbox{sign}}
\newcommand{\epo}{\,.}

\newcommand{\bit}{\begin{itemize}}
\newcommand{\eit}{\end{itemize}}
\begin{document}

\title{
\vskip-3cm{\baselineskip14pt
\centerline{\normalsize DESY~14-092,~~HU-EP-14/25\hfill}
\centerline{\normalsize June 2014\hfill}}
\vskip1.5cm
About the role of the Higgs boson in the evolution of the early universe
\footnote{Presented at the Cracow Epiphany Conference on physics at the
LHC, Krak\`ow, Poland, January 8-10, 2014.}
}

\author{
{\sc Fred Jegerlehner},
\\
\\
{\normalsize Humboldt-Universit\"at zu Berlin, Institut f\"ur Physik,}\\
{\normalsize  Newtonstrasse 15, D-12489 Berlin, Germany}\\
{\normalsize Deutsches Elektronen-Synchrotron (DESY),}\\
{\normalsize Platanenallee 6, D-15738 Zeuthen, Germany}
}

\date{}

\maketitle
\begin{abstract}{\footnotesize
I review a recent analysis presented
in~\cite{Jegerlehner:2013cta,Jegerlehner:2013nna,Jegerlehner:2014mua}.
After the discovery of the Higgs particle the most relevant structures
of the SM have been verified and for the first time we know all
parameters of the SM within remarkable accuracy. Together with recent
calculations of the SM renormalization group coefficients up to three
loops we can safely extrapolate running couplings high up in
energy. Assuming that the SM is a low energy effective theory of a
cutoff theory residing at the Planck scale, we are able to calculate
the effective bare parameters of the underlying cutoff system. It
turns out that the effective bare mass term changes sign not far below
the Planck scale, which means that in the early universe the SM was in
the symmetric phase. The sign-flip, which is a result of a conspiracy
between the SM couplings and their screening/antiscreening behavior,
triggers the Higgs mechanism. Above the Higgs phase transition the
bare mass term in the Higgs potential must have had a large positive
value, enhanced by the quadratic divergence of the bare Higgs
mass. Likewise the quartically enhanced positive vacuum energy density
is present in the symmetric phase. The Higgs system thus provides the
large dark energy density in the early universe, which triggers
slow-roll inflation, i.e. the SM Higgs is the inflaton scalar
field. Reheating is dominated by the decay of the heavy Higgses into
(in the symmetric phase) massless top/anti-top quark pairs.  The Higgs
mechanism stops inflation and the subsequent electroweak phase
transition provides the masses to the SM particles in proportion to
their coupling strength. The previously most abundantly produced
particles are now the heaviest and decay into the lighter ones, by
cascading down the CKM-element matrix from top and bottom to normal
matter.  Baryon-number $B$ violating interactions are naturally
provided by Weinberg's set of close-by dimension 6 four-fermion
effective interactions.  Since matter is produced originating from the
primordial heavy Higgs fields via $C$ and $CP$ violating decays we
have actually a new scenario which possibly could explain the
baryon-asymmetry essentially in terms of SM physics.}
%
\end{abstract}

\renewcommand{\thefootnote}{\arabic{footnote}}
\setcounter{footnote}{0}

\section{Introduction}
With the discovery of the Higgs boson by ATLAS~\cite{ATLAS} and CMS~\cite{CMS} at the LHC all
relevant ingredients of the Standard Model (SM) have been established
experimentally. In particular, for the first time we know all the
basic SM parameters with remarkable accuracy. The Higgs mass, found to
be $M_H=125.9\pm0.4~\gv$, turned out to have a value just in the
window which was required to allow one to extrapolate SM physics up to
the Planck scale~\cite{Hambye:1996wb} without the need to assume some
new non-SM physics. This together with the fact that so far no hints
for a supersymmetric extension or extra dimensions etc. have been
found, sheds new light on the structure of the SM and its
self-consistency.  The SM together with its specific values for the
couplings, the gauge couplings $g'$, $g$, $g_s$, the top-quark Yukawa
coupling $y_t$ and newly, the Higgs self-coupling $\lambda$ are
supporting the picture of the SM as a low energy effective theory of
some cutoff system residing at the Planck scale. In such a framework
the relation between bare and renormalized physical low energy
parameters acquirers a physical meaning and from the knowledge of the
physical parameters we can calculate actually the bare parameters
relevant at the high (short distance) scale. The SM as a low energy
theory, is then emerging as a result of the low energy expansion in
$E/\Lambda_{\rm Pl}$.  All positive powers $\left(E/\Lambda_{\rm
Pl}\right)^n$, $n=1,2,3,
\cdots$ are heavily suppressed by the very high cutoff $\Lambda_{\rm
Pl}\sim 10^{19}~\gv$ and unobservable at present accelerator
energies. Renormalizability of the SM as well as all known conditions
which where required to get the SM as a minimal renormalizable
extension of its low energy effective structure now are a consequence
of the low energy expansion. As we do not see the infinite tower of
non-renormalizable effective operators, the low energy effective
theory actually has more symmetry than the underlying cutoff system at
the Planck scale, which is largely unknown in its details.  In such a
scenario simplicity and symmetries are expected to be naturally
generated dynamically as a consequence of our blindness for the
details of the underlying cutoff system.

Our scenario, not new at all, has to be seen in the context of the
general question about ``What is the path to physics at the Planck
scale?''. The \textit{String Paradigm} assumes that ``the closer we
look the more symmetric the world looks like'' assuming a hierarchy
of symmetries like
\begin{center}
{ M--T{\small
heory} $\sim$ Strings} { $\lear$ SUGRA $\lear$ SUSY
$\lear$ SM},
\end{center}
In contrast the \textit{Emergence Paradigm} understands nature as
``the less close you look the simpler it looks''
\begin{center}
Planck medium $\sim$ {``E{\small ther''} $\to$ low energy effective QFT $\to$
SM}.
\end{center}
The latter view understands the SM as the ``true world'' seen from far
away. The methodological approach we know from investigating the long
range properties of condensed matter systems, specifically,
critical phenomena, which may be applied to particle physics as
well. In this context even the quantum field theories are structures
emergent from critical and quasi critical underlying condensed matter
systems.

At the high scale, given by the intrinsic cutoff, one expects all kinds
of excitations. Most of them cannot be seen at long distances
(non-critical modes), however, conspiracies between modes are able to
develop quasi critical modes which are seen as light particles in
interaction, which take the form of a non-trivial renormalizable QFT in
space-time of dimension $D=4$. For $D>4$ only trivial stable theories
would exist, such that extra dimensions decouple.  About the details
of the ``ether'' we do not know much, except that we have to stay
within the universality class ($\equiv$ the totality of possible
systems exhibiting identical long range tail) of the SM. Such a view
turns upside-down the standard believes that the higher the energy the
simpler the world, together with the assumption that symmetries in
nature are broken at best spontaneously. Such scenarios assume
renormalizability as a basic principle and symmetries to be broken by
the relevant operators of dimension $d<4$ and ignore the fact
that there is a infinite tower of possible higher order operators with $d>4$,
which generally would violate symmetries seen at low energy.

Symmetries relevant for the SM are small gauge groups, with particles
in multiplets of few conspiring fields, like doublets and triplets,
i.e.  the SM gauge structure is natural in a low energy expansion.
In contrast, GUT symmetries are not naturally emergent and have
to be put by hand at the high scale.

In this scenario the relation between bare and renormalized parameters
is physical and bare parameters predictable from known renormalized
ones. All so called ``UV singularities'' must be taken serious
including terms enhanced quadratically and quartically in the
cutoff. Since the cutoff is finite there are no divergences and a
cutoff limit is not required to exist. The impact of the very high
Planck cutoff is that the local renormalizable QFT structure of the SM
is presumably valid up not far below the Planck scale. This also
justifies the application of the SM RG up to high scales.

\section{Low energy effective QFT of a cutoff system}
I think it is instructive to be more specific about the appearance of
low energy effective theories. The best would be to implement the SM as
a lattice field theory in the unitary gauge, in which the Higgs vacuum
expectation value (VEV)
$v$ is a well defined order parameter of the broken
$H\leftrightarrow-H$ ($Z_2$) symmetry, with a small lattice spacing
$a=\Lambda^{-1}$, and take the lattice system as the true underlying
theory and work out its long range properties. In order to illustrate
the emergence of a low energy effective theory, for simplicity, let us
consider the cutoff version of a self-interacting Higgs system with
Lagrangian (for details see ~\cite{Jegerlehner:1976xd})
\bea
\cL =\cL_0+\cL_{\rm int}=\frac12
\,\partial^\mu \phi(x)\,(1+\Box/\Lambda^2)\,\partial_\mu \phi(x)
-\frac12\,m_0^2\,\phi(x)^2-\frac{\lambda_0}{4!}\,\phi^4(x)\epo
\eea
The regularization is chosen here as a Pais-Uhlenbeck
higher-derivative kinetic cutoff term. 

We consider a vertex function (connected amputated one-particle irreducible diagrams) of
$N$ scalar fields. The bare vertex functions are related to the renormalized
ones by reparametrizing parameters and fields
\bea
\Gamma^{(N)}_{\Lambda\,r}(p;m,\lambda)=Z^{N/2}(\Lambda/m,\lambda)\,
\Gamma^{(N)}_{\Lambda\,b}(p;\Delta m_0(\Lambda,m,\lambda),\lambda_0(\Lambda/m,\lambda))\epo
\eea
The renormalized functions satisfy a RG equation which controls the response to a
change of the cutoff $\Lambda$: 
$\left.\Lambda \frac{\partial}{\partial
\Lambda}\,\Gamma{(N)}_{\Lambda\,b}\,\right|_{m,\lambda}$, for fixed
renormalized parameters, and which by applying
the chain rule of differentiation yields
\bea
&&\left(\Lambda \frac{\partial}{\partial \Lambda}+\beta_0
\frac{\partial}{\partial \lambda} -N\,\gamma_0+ \delta_0\, \Delta
m_0^2\,\frac{\partial}{\partial \Delta
m_0^2}\right)\,\Gamma^{(N)}_{\Lambda\,b}(p;m_0,\lambda_0)\nn \\ &=&
Z^{-N/2}\,\Lambda \frac{\partial}{\partial \Lambda}\,\Gamma^{(N)}_{\Lambda\,r}(p;m,\lambda)\epo
\label{LamRG}
\eea
$m^2_{0c}$ is the ``critical value'' of the bare mass for which the
renormalized mass is zero, i.e.
$\left.\Gamma^{(2)}_{\Lambda\,b}\right|_{p=0}=0$, and $\Delta
m_0^2=m_0^2-m_{0c}^2$ corresponds to the renormalized mass parameter.
Since the renormalized vertex functions have a regular limit as
$\Lambda \to \infty$, to all orders in perturbation theory the
inhomogeneous part behaves as
\bea
Z^{N/2}\,\Lambda \frac{\partial}{\partial \Lambda}\,\Gamma^{(N)}_{\Lambda\,r}
(p; m,\lambda)=O(\Lambda^{-2}(\ln \Lambda)^l)\;,
\eea
i.e., the inhomogeneous part, representing a cutoff insertion, falls
off faster than the l.h.s. of Eq.~(\ref{LamRG}) by two powers in the
cutoff for large cutoffs. This is easy to understand given the fact
that the cutoff enters $\cL$ as a term proportional to
$\Lambda^{-2}$. All the RG equation
coefficients exist as non-trivial functions in the limit of infinite cutoff:
\bea
\lim_{\Lambda \to \infty} \alpha_0(\Lambda/m,\lambda) =\alpha(\lambda)\,,\:\alpha=\beta,\gamma,\delta\,,
\eea
for dimensions $2\leq D \leq 4$. In $D=4$ dimensions the
proper vertex-functions have a large cutoff $\Lambda$-expansion 
\bea
\Gamma^{(N)}_{\Lambda\,b}(p;\Delta m_0,\lambda_0)=\sum_{k,l \geq
0} \Lambda^{-2k}(\ln \Lambda)^l\,f^{(N)}_{kl}(p\;\Delta m_0, \lambda_0)\,,
\eea
and for large $\Lambda$ we obtain the \underline{preasymptote} of $\Gamma^{(N)}_{\Lambda\,b}$
\bea
\Gamma^{(N)}_{\Lambda\,{\rm as}}(p;\Delta m_0,\lambda_0)=\sum_{l \geq
0} (\ln \Lambda)^l\,f^{(N)}_{0l}(p\;\Delta m_0, \lambda_0\Lambda^\veps)\,,
\eea
which is collecting all leading terms and satisfies the bound
\bea
\left|\Gamma^{(N)}_{\Lambda\,b}(p;\Delta
m_0,\lambda_0)-\Gamma^{(N)}_{\Lambda\,{\rm as}}(p;\Delta m_0,\lambda_0) \right|=O(\Lambda^{-2}(\ln \Lambda^{l_x}))\epo
\eea
The index $l_x$ is bounded to all orders in the perturbation
expansion. The key point is that the still cutoff dependent preasymptote satisfies a
\underline{homogeneous} RG equation, a special property of the long range
tail of the bare theory:
\bea
&&\left(\Lambda \frac{\partial}{\partial \Lambda}+\beta_{\rm
as}(\Lambda/\Delta m_0,\lambda_0)
\frac{\partial}{\partial \lambda_0} -N\,\gamma_{\rm
as}(\Lambda/\Delta m_0,\lambda_0)\right. \nn \\ && \left. \hspace*{1cm}
+ \delta_{\rm
as}(\Lambda/\Delta m_0,\lambda_0)\, \Delta
m_0^2\,\frac{\partial}{\partial \Delta
m_0^2}\right)\,\Gamma^{(N)}_{\Lambda\,{\rm as}}(p;\Delta m_0,\lambda_0)=0\epo
\label{preasRG}
\eea
The homogeneity of this partial differential equation for the response
to a change in $\Lambda$ means that $\Lambda$ does not represent a cutoff any more
and just takes the role of a renormalization scale parameter. The interpretation
(verifiable to all orders in the perturbation expansion) is the following:
\bit
\item the preasymptotic theory is a non-trivial local relativistic QFT;

\item the crucial point is that the cutoff $\Lambda$ is physical i.e. a finite number
and by a finite renormalization (renormalizing parameters and fields
only) by change of scale $p_i \to \kappa\,p_i\semis
\kappa=\Lambda/\mu$ one can achieve that momenta measured in units of
$\Lambda$ are rescaled to momenta expressed in units of \MSb scale $\mu$;

\item as a consequence, the relationship between renormalized and bare parameters is
      physical, such that knowing the renormalized parameters we are
      able to calculate the bare ones;

\item an important empirical fact: besides QCD at low energy, elementary particle
      interactions have rather weak coupling such that perturbation theory works in general;

\item applied to our ``real world'' physics with $\Lambda=\LPl$ the cutoff
      is very very high such that all cutoff structure are deeply
      hidden at present accelerator energies.
\eit

A comparison with QCD reveals the importance of a large cutoff.
Low energy effective hadron theories suffer from the close-by cutoff
and are therefore difficult to establish unambiguously. 
In Table~\ref{tab:LEE} we give another representation of the low
energy expansion at work:
\begin{table}[h]
\centering
\caption{Typical operators in a low energy expansion}
\label{tab:LEE}
\begin{tabular}{ccccc}
\hline
\hline
 & dimension & operator & scaling behavior & \\
\hline
&&&\\[-3mm]
&$\cdot$&$\infty$--many&&\\
&$\cdot$&irrelevant &&\\
$\uparrow$&$\cdot$&operators&&\\
no&&&&\\
data & $d=6$ & $ (\Box \phi)^2, (\bar{\psi}\psi)^2, \cdots $&
 $ (E/\LPl)^2$& \\
$|$ & $d=5$ & $ \bar{\psi}\sigma^{\mu\nu}F_{\mu\nu}\psi, \cdots $&
 $ (E/\LPl)$ &\\
&&&&\\
\hline
&&&&\\
\multirow{6}{*}{\begin{tabular}{c}$|$\\
experimental \\ data \\ $\downarrow$\end{tabular}} & $d=4$ & $(\partial \phi)^2, \phi^4, (F_{\mu\nu})^2, \cdots $&
 $ \ln (E/\LPl)$ \\
&&&&\\
\cline{2-5}
&&&&\\
 & $d=3$ & $\phi^3, \bar{\psi}\psi $&
 $ (\LPl/E)$ & 
\\
 & $d=2$ & $ \phi^2, (A_\mu)^2 $&
 $ (\LPl/E)^2$ & \\
 & $d=1$ & $\phi$&
 $ (\LPl/E)^3$&\\
&&&&\\
\end{tabular}
\end{table}

The relevant operators must be tamed by symmetries, in order not to
blow up with the cutoff: chiral symmetry and gauge symmetry in the SM,
and supersymmetry in supersymmetric extensions of the SM.

Up to date and for a long time to come there is and will be no direct
experimental information on $O(E/\LPl)$ or $O(E^2/\LPl^2)$ effects,
but bounds on the absence of such terms, unless they violate basic SM
symmetries like baryon-number conservation, for example.

The infinite tower of \textbf{irrelevant operators} of dimension $>4$
are not seen at low energy and imply the simplicity of the SM! 
Blindness to details implies more symmetries (Yang-Mills structure
[gauge cancellations] with small groups\footnote{Such a pattern (few
particle multiplets) reminds of primordial nucleosynthesis, which
exclusively produces only the simplest, i.e. lightest, elements.}
: doublets, triplets besides
singlets, Lorentz invariance, anomaly cancellation and family
structure, triviality for space-time dimensions $D>4$ [D=4 border
case for an interacting world at long distances, this has nothing to do with
compactification, extra dimensions just trivialize by themselves],
etc.). The natural emergence of spin 1 and spin 2 excitations has been
considered in Ref.~\cite{Jegerlehner:1978nk}.

Problems are posing the \textbf{relevant operators} of dimension
$<4$. In particular the mass terms, require ``tuning to
criticality''. In the symmetric phase of the SM we are confronted with
one mass term only (the others are forbidden by the known chiral and
gauge symmetries), the one of the Higgs doublet field.

The symmetric phase Higgs fine tuning has the form
\bea
m_0^2=m^2+\delta m^2\;;\;\; \delta m^2= \frac{\Lambda^2}{32 \pi^2}\,C\,,
\label{Higgsmassreno}
\eea
with a coefficient typically $C=O(1)$. To keep the renormalized mass
at some small value, which can be seen at low energy, $m^2_0$ has to
be adjusted to compensate the huge number $\delta m^2$ such that about
\textbf{35 digits} must be adjusted in order to get the observed value around
the electroweak scale. This is the usual hierarchy problem.
  
\section{Matching and running couplings}
The key questions asked here are: 1) how does SM physics look like at
much higher energies and 2) what does the Higgs potential look like at
the bare level, when going to the Planck scale. The first question can
be answered, under the assumption that no substantial effects come in by
possible physics beyond the SM, by studying the evolution of couplings
as determined by the SM renormalization group (RG), which now is known
to three loops in the \MSb renormalization
scheme~\cite{Mihaila:2012fm,Bednyakov:2012rb,Bednyakov:2012en,Bednyakov:2013eba,Chetyrkin:2012rz,Chetyrkin:2013wya}.
The initial \MSb values have to be obtained by appropriate matching
conditions from the physical on-shell parameters. For the latter we
use the values~\cite{pdg}:
{\small
\begin{eqnarray}
&&
M_Z = 91.1876(21)~\gv,
\quad
M_W = 80.385(15)~\gv,
\quad
M_t = 173.5(1.0)~\gv,
\nonumber \\ &&
G_\mu =1.16637(1)\times 10^{-5}~\gv^{-2}
\;,\;\;\hat{G}_\mu =G_\mu(M_Z)=1.15564(55)\times 10^{-5}~\gv^{-2}
\nonumber \\ &&
\alpha^{-1} = 137.035999\,,\;\;\alpha^{-1}(M_Z^2) = 127.944\,,\;\;
\alpha_s(M_Z^2) = 0.1184(7)\;.
\label{params}
\end{eqnarray}
For the Higgs mass we adopt
\begin{eqnarray}
M_H = 125.9\pm 0.4~\gv,
\label{mhinp}
\end{eqnarray}
}
\noindent
in accord with latest ATLAS and CMS reports. All light-fermion masses
$M_f\,(f\neq t)$ give negligible effects and do not play any role in
our consideration. The top quark mass given above is taken to be the
pole mass. It should be reminded that it is not precisely clear
whether the value reported by experiments or by the PDG can be
identified with the on-shell mass within the given accuracy. For a
recent review on the subtleties in defining/measuring the top quark
mass see e.g.~\cite{Juste:2013dsa} and
references therein.

One somewhat controversial issue about the electroweak matching
conditions concern the inclusion of tadpole contributions in the
relationship between on-shell and \MSb quantities. The tadpoles which
only show up in the broken phase, where they results from the
radiative corrections of the Higgs VEV $v$, on the one hand can yield
large corrections, on the other hand there is a theorem which says
that tadpole contributions drop out from relations between measured
quantities. For this reason tadpoles are often dropped in actual
calculations. It should be realized that experimentally measured
quantities incorporate tadpole contributions in any case (one cannot
exclude subsets of diagrams form a measurement). The relation between
\MSb and on-shell quantities, however is not a relation between
physical quantities and tadpoles are relevant to be included. Since
tadpoles are neither gauge-invariant nor UV finite, dropping them
leads to gauge dependent quasi-\MSb parameters which in addition do
not satisfy the correct RG equations
(see~\cite{Fleischer:1980ub,Jegerlehner:2001fb,Jegerlehner:2003py,Jegerlehner:2012kn}).
Care has to be taken also of the fact that the weak corrections are
not respecting the Appelquist-Carazzone
theorem~\cite{Appelquist:1974tg} when evaluating the matching
conditions. This means for example that electroweak top quark
contributions do not start above the top quark threshold. Top quarks
e.g. give a large contribution to the $\rho$-parameter
\bea
\rho(0) &\equiv&
G_{\rm NC}(0)/G_\mu(0)=1+\frac{3\sqrt{2}G_\mu}{16\pi^2}\,\biggl\{M_t^2
\crn &&+\left(\frac{M_W^2}{1-M_W^2/M_H^2}\ln
M_H^2/M_W^2-\frac{M_Z^2}{1-M_Z^2/M_H^2}\ln
M_H^2/M_Z^2+\cdots\right)\biggr\}\,,\nn 
\eea
where $G_{\rm NC}(0)$ and
$G_\mu(0)$ are the neutral and charged current effective Fermi
couplings at zero momentum, respectively.

The top Yukawa coupling and the Higgs self-coupling are only known via
their measured masses via the mass coupling relations
\begin{eqnarray}
m_W^2(\mu^2)&=&\frac14\,g^2(\mu^2)\,v^2(\mu^2)\semis
m_Z^2(\mu^2)=\frac14\,(g^2(\mu^2)+g'^2(\mu^2))\,v^2(\mu^2)\semis\crn
m_f^2(\mu^2)&=&\frac12\,y^2_f(\mu^2)\,v^2(\mu^2)\semis
m_H^2(\mu^2)=\frac13\,\lambda(\mu^2)\,v^2(\mu^2)\comas
\label{vevsquare}
\end{eqnarray}
which derive from the Higgs mechanism.

\begin{figure}[t]
\centering
\includegraphics[height=5.8cm]{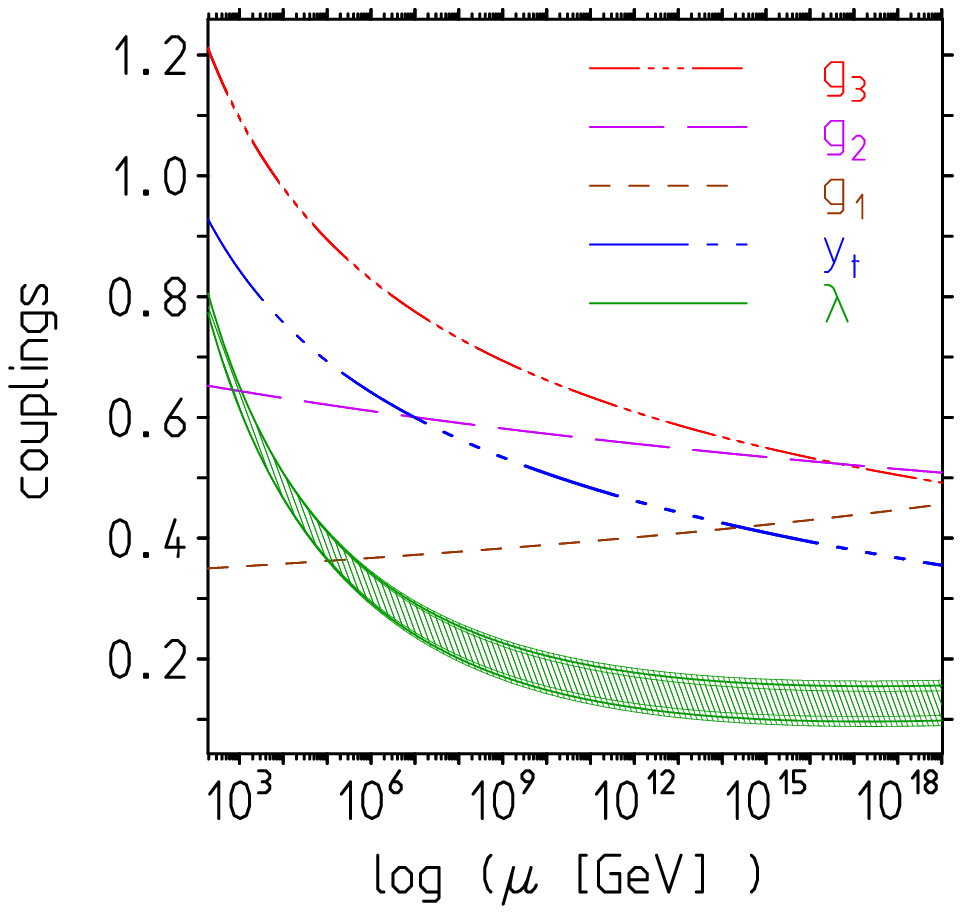}
\includegraphics[height=5.8cm]{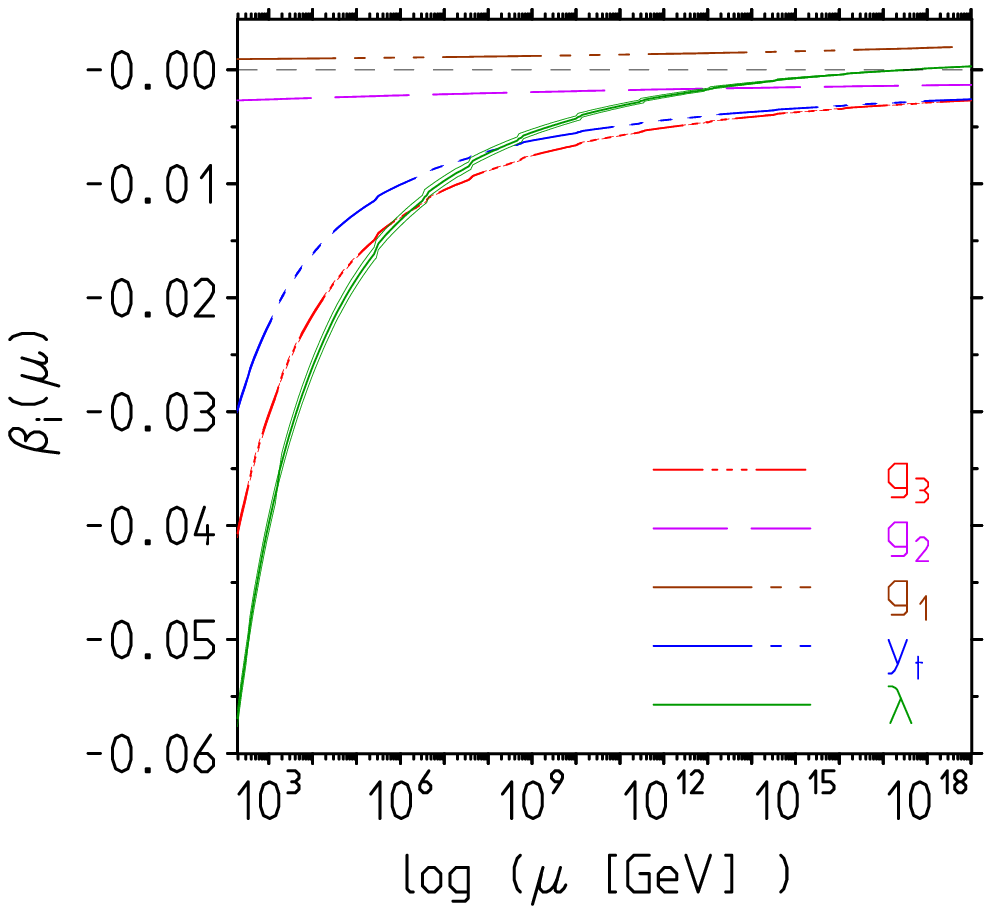}
\caption{Left: the dimensionless SM couplings in the \MSb scheme as a
function of the renormalization scale (see also~\cite{Yukawa:3,degrassi,Masina:2012tz,Hamada:2012bp,Buttazzo:2013uya}). 
The input parameter uncertainties as given in RPP~\cite{pdg} are
represented by the thickness of the lines. The gray/green band
corresponds to Higgs masses in the range [124-127]~GeV. Right: the
$\beta$-functions for the couplings $g_3$, $g_2$, $g_1$, $y_t$ and $\lambda$.
The uncertainties are represented by the line widths.}
\label{fig:SMrunpar}
\end{figure}

We will have to distinguish bare and renormalized quantities and among
the latter \MSb and physical on-shell ones. As usual we adopt
dimensional renormalization starting form $D=4-\veps$ dimensions and
taking the limit $\veps \to +0$ after renormalization. By $m_{i0}$ we
denoted the bare, by $m_i$ the \MSb and by $M_i$ the on-shell
masses. $\Reg=\frac{2}{\varepsilon}-\gamma +
\ln 4 \pi+\ln \mu_{0}^2$ is the UV regulator term with $\mu_{0}$ the
bare scale parameters used in dimensional renormalization.
The substitution $\Reg \to \ln \mu^2$ defines the UV finite
\MSb parametrization. Let $\delta M_{b}^2$ denote the bare on-shell
mass counterterm for a boson species $b$ and $\delta M_{f}$ the
corresponding counterterm for a fermion species $f$. By identifying
$m_b^2(\mu^2)=M_{b}^2+\delta M^2_b|_{\Reg =\ln \mu^2}$ and $m_{f}(\mu^2)=M_{f}
+\delta M_f|_{\Reg =\ln
\mu^2}$, respectively, we then obtain the \MSb masses in terms of the
on-shell masses.  Similar relations apply for the coupling constants
$g$, $g'$, $\lambda$ and $y_f$, which, however, usually are fixed
using the mass-coupling relations in terms of the masses and the Higgs
VEV, which is determined by the Fermi constant as
$v=(\sqrt{2}G_\mu)^{-1/2}$. Here $G_\mu$ is the muon decay constant,
which represents the Fermi constant in the on-shell scheme. The \MSb
version of the Fermi constant we denote by $G_{F}^{\overline{\rm MS}}$
or simply by $G_{F}$. The matching condition for the Higgs VEV may be
represented in terms of the one for the muon decay
constant
\bea
G_{F}^{\overline{\rm MS}}(\mu^2)=G_\mu+\left(\left.\delta G_\mu \right|_{\rm OS}\right)_{\Reg =\ln \mu^2}\,,
\eea
where $\left.\frac{\delta G_\mu}{G_\mu}\right|_{\rm
OS}=2\,\frac{\delta v^{-1}}{v^{-1}}\,.$ for details I refer to
Ref.~\cite{Fleischer:1980ub,Jegerlehner:2012kn}. Then the \MSb top
quark Yukawa coupling is given by
\bea
y_t^{\overline{\rm MS}}(M_t^2)=
\sqrt{2}\,\frac{m_t(M_t^2)}{v^{\overline{\rm MS}}(M_t^2)}\semis v^{\overline{\rm MS}}(\mu^2)
=\left(\sqrt{2}\,G_{F}^{\overline{\rm MS}}\right)^{-1/2}(\mu^2)\,,
\eea
and the other \MSb mass-coupling relations correspondingly.
The RG equation for $v^2(\mu^2)$ follows from the RG equations
for masses and coupling of the Higgs potential $V(\phi)=\frac12\,m^2\,\phi^2+\frac{1}{24}\,\lambda\,\phi^4$
as
\begin{eqnarray}
 v^2(\mu^2)=3\,\frac{m_H^2(\mu^2)}{\lambda(\mu^2)}\semis
\mu^2 \frac{d}{d \mu^2} v^2(\mu^2)
=
v^2(\mu^2) \left[\gamma_{m^2}  - \frac{\beta_\lambda}{\lambda} \right]\epo
\label{vev}
\end{eqnarray}
We remind that all dimensionless couplings satisfy the same RG
equations in the broken and in the unbroken
phase. Figure~\ref{fig:SMrunpar} shows the solutions of the RG
equations and the $\beta$-functions up to $\mu=M_{\rm Planck}$.

Remarkably, as previously found for the running couplings in
Refs.~\cite{Yukawa:3,degrassi,Masina:2012tz,Hamada:2012bp,Buttazzo:2013uya}, all
parameters stay in bounded ranges up to the Planck scale if one adopts
our matching conditions together with the so far calculated RG
coefficients. We note that including all known terms no transition to
a metastable state in the effective Higgs potential is observed with
our set of \MSb input parameters,
i.e. no change of sign in $\lambda$ occurs, in agreement with
Refs.~\cite{Yukawa:3,Masina:2012tz}. Results at various scales are
collected in Table~\ref{tab:params}.
\begin{table}[h]
\caption{Parameters in \MSb scheme at various scales for $M_H=126~\gv$ and
$\mu_0\simeq 1.4\power{16}~\gv$. $C_1$ and $C_2$ are the one- and
two-loop coefficients of the quadratic divergence,
respectively. $C_1$ given by Eq.~(\ref{coefC1}). The last two columns show corresponding results from
Ref.~\cite{Buttazzo:2013uya}.}  {\begin{tabular}{ccccc||cc}
\hline\noalign{\smallskip}
coupling $\backslash$ scale  & $M_Z$ & $M_t$ & $\mu_0$ & $\mpl$ & $M_t$~\cite{Buttazzo:2013uya} &
$\mpl$~\cite{Buttazzo:2013uya} \\ 
\hline\noalign{\smallskip}
$g_3$ &   $1.2200$ & $1.1644$ & $0.5271$ & $0.4886$ & 1.1644 &\hphantom{-} 0.4873 \\
$g_2$ &   $0.6530$ & $0.6496$ & $0.5249$ & $0.5068$ & 0.6483 &\hphantom{-} 0.5057 \\
$g_1$ &   $0.3497$ & $0.3509$ & $0.4333$ & $0.4589$ & 0.3587 &\hphantom{-} 0.4777 \\
$y_t$ &   $0.9347$ & $0.9002$ & $0.3872$ & $0.3510$ & 0.9399 &\hphantom{-} 0.3823 \\
$y_b$ &   $0.0238$ & $0.0227$ & $0.0082$ & $0.0074$ && \\
$y_\tau$ &$0.0104$ & $0.0104$ & $0.0097$ & $0.0094$ && \\
$\sqrt{\lambda}$&$0.8983$ & $0.8586$ & $0.3732$ & $0.3749$ & 0.8733  & $\I\:\,$0.1131 \\
$\lambda       $&$0.8070$ & $0.7373$ & $0.1393$ & $0.1405$ & 0.7626  &
- 0.0128         \\
$C_1$ & $-6.768$ & $-6.110$ & $\hphantom{0}0\hphantom{.0000}$ & $0.2741$ & $$ & $$\\
$C_2$ & $-6.672$ & $-6.217$ & $\hphantom{0}0\hphantom{.0000}$ & $0.2845$ & $$ & $$\\
$m[GeV]$&$89.096$ & $89.889$ & $97.278$ & $96.498$ &97.278&  \\ 
\hline\noalign{\smallskip}
\end{tabular}
\label{tab:params}}
\end{table}

\section{The quadratic divergences in the SM}
In the unbroken phase the only quadratic divergences show up in the
renormalization of Higgs potential mass $m$. Since the UV structure is the same in the
broken phase, there are no other problems in this direction.
Here we encounter the fine tuning relation (\ref{Higgsmassreno}).
At one-loop the coefficient function $C_1$ 
has been discussed within this context by Veltman~\cite{Veltman:1980mj},
and modulo small lighter fermion
contributions is given by
\bea
C_1={ \frac{6}{v^2}(M_H^2 + M_Z^2
+2 M_W^2-4 M_t^2)}=2\,\lambda+\frac32\, {g'}^{2}+\frac92\,g^2-12\,y_t^2\epo
\label{coefC1}
\eea
On the one hand parameters are known in the broken low energy phase,
where they are directly accessible to experiment, on the other hand
they are given in terms of SM parameters in the unbroken phase, which
is physical at high energies. A priori, the renormalized $m^2$ in the
symmetric phase is not known and not accessible directly to
experiment. As we will see below, if $m^2$ would not be small relative
to the very large $\delta m^2$ it would affect the inflation pattern
and thus in principle is constrained by the observed Cosmic Microwave
Background (CMB) fluctuation data. In fact the matching condition
$m^2_0=m^2$ at scale $\mu_0$ where $\delta
m^2=0$ actually fixes the renormalized mass at any scale in terms of
the measured Higgs mass and the RG evolution of it. So the hierarchy
problem seems to be a problem in the symmetric phase. In order to
understand this we have to be aware that the Higgs is not a fundamental
mode in the underlying cutoff system.  Therefore, in the underlying
cutoff system $m_0^2$ is not a fundamental parameter, but an effective
one associated with the scalar Higgs mode, which usually is some
collective effect within the Planck medium. This means that the
effective bare mass is actually essentially generated by the dynamics
and hence largely determined by $\delta m^2$, i.e. the bare mass is
radiatively generated. In any case we assume $m^2$ to be small
relative to $\delta m^2$.

\begin{figure}
\centering
\includegraphics[height=5.8cm]{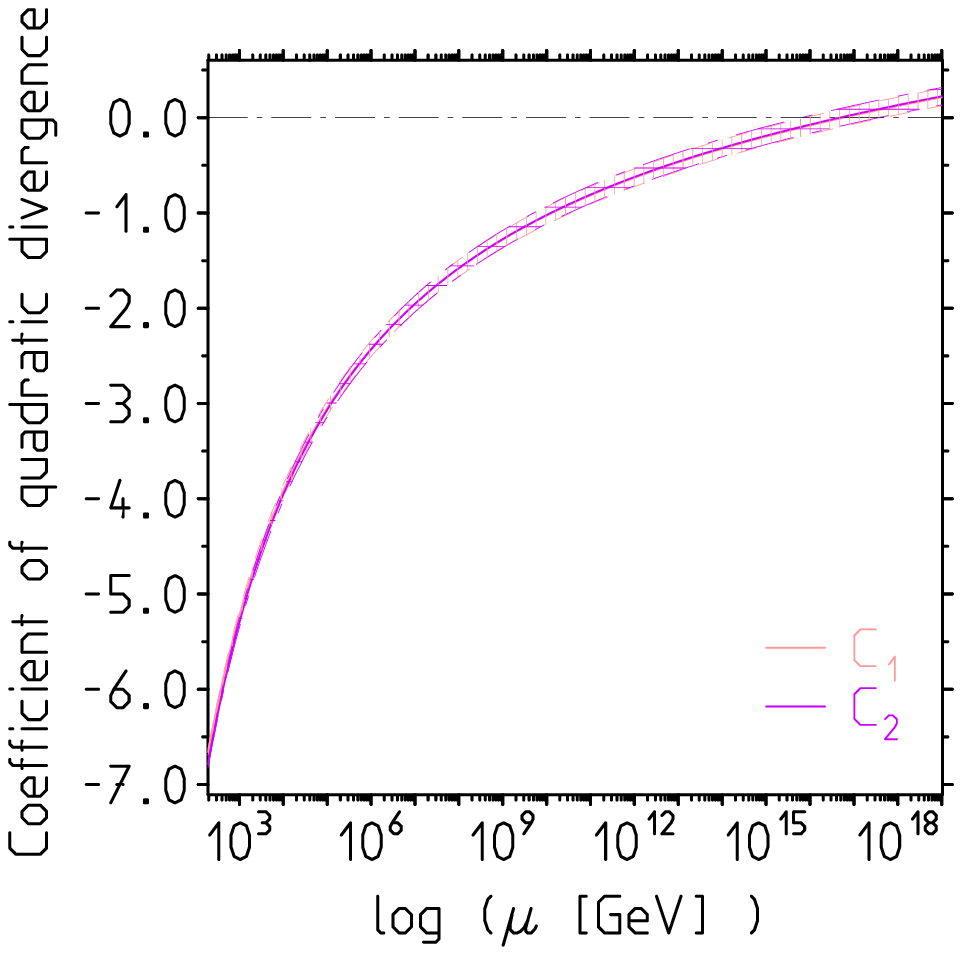}
\includegraphics[height=5.8cm]{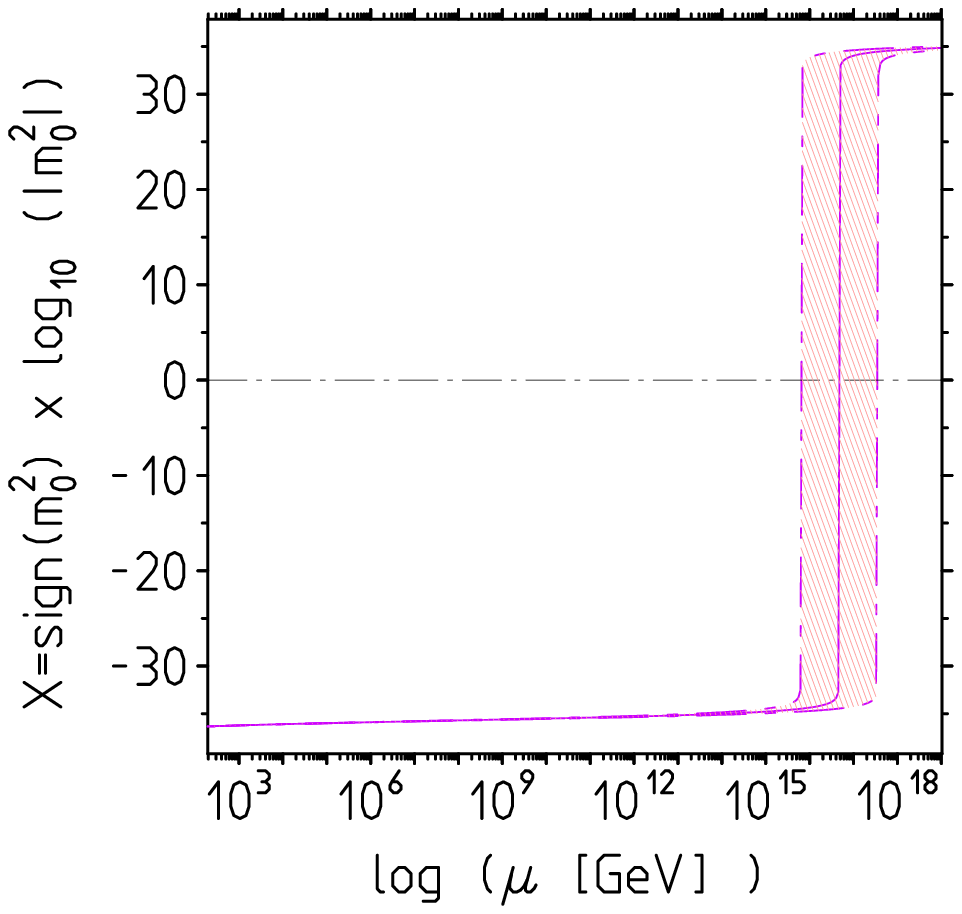}
\caption{The Higgs phase transition in the SM. Left: the zero in
$C_1$ and $C_2$ for $M_H=125.9\pm 0.4~\gv$. Right: shown is $
X=\sign(m^2_0)\times \log_{10} (|m^2_0|)$, which represents
$ m^2_0=\sign(m^2_0)
\times 10^{X}$.}
\label{fig:quaddivcoef}
\end{figure}

What is important is that $C_1$ is universal and depends on dimensionless
gauge, Yukawa and Higgs self-coupling only, the RGs of which are
unambiguous. Similarly, for the two-loop coefficient $C_2$, first calculated
in Refs.~\cite{Alsarhi:1991ji,Jones:2013aua},
\bea
C_2&=&C_1+ \frac{\ln (2^6/3^3)}{16\pi^2}\, [
18\,y_t^4+y_t^2\,(-\frac{7}{6}\,{g'}^2+\frac{9}{2}\,g^2
             -32\,g_s^2) \nn \\
             &&-\frac{87}{8}\,{g'}^4-\frac{63}{8}\,g^4 -\frac{15}{4}\,g^2{g'}^2
             +\lambda\,(-6\,y_t^2+{g'}^2+3\,g^2)
             -\frac{2}{3}\,\lambda^2]\,,
\label{coefC2}
\eea
which numerically does not change significantly the one-loop
result. Recently, Hamada, Kawai and Oda~\cite{Hamada:2012bp} have
investigated the coefficients to two loops in terms of running
couplings and found the coefficients of the quadratic divergence to
have a zero not far above the Planck scale.  For the parameters listed
in Table~\ref{tab:params}, the SM makes a prediction for the
coefficients $C_i$ and hence for the bare mass parameter in the Higgs
potential, which we displayed in Fig.~\ref{fig:quaddivcoef}.
In the broken phase given by $m^2_0=\frac12\,m^2_{H0}$, $m^2_0$ is calculable and is exhibiting the following
properties: i) the coefficient $C_n(\mu)$ exhibits a zero, for
$M_H=126~\gv$ at about $\mu_0\sim 1.4 \power{16}$, not far below $
\mu=M_{\rm Planck}$, ii) at the zero of the coefficient function the
counterterm $ \delta m^2=m^2_0-m^2=0$
vanishes and the bare mass changes sign, iii) this represents a first
order phase transition which triggers the Higgs mechanism and seems to
play an important role for
\textit{cosmic inflation}, iv) at the transition point $\mu_0$ we have
$v_0=v(\mu_0^2)$, where $ v(\mu^2)$ is the \MSb renormalized
Higgs VEV, v) the jump in the vacuum density, thus agrees with the
renormalized one: $ -\Delta \rho_{\rm
vac}=\frac{\lambda(\mu^2_0)}{24}\, v^4(\mu_0^2)\,,$ and thus is $
O(v^4)$ and \textbf{not} $ O(M^4_{\rm Planck})\,.$

We note that $ \beta_\lambda$ has a zero at about $
\mu_\lambda \sim 3.5 \power{17} > \mu_0$, where the Higgs self-coupling $
\lambda$ although rather small is still positive and then starts
slowly increasing up to $M_{\rm Planck}$~\cite{Jegerlehner:2012kn}.

\begin{figure}
\centering
\includegraphics[height=5.8cm]{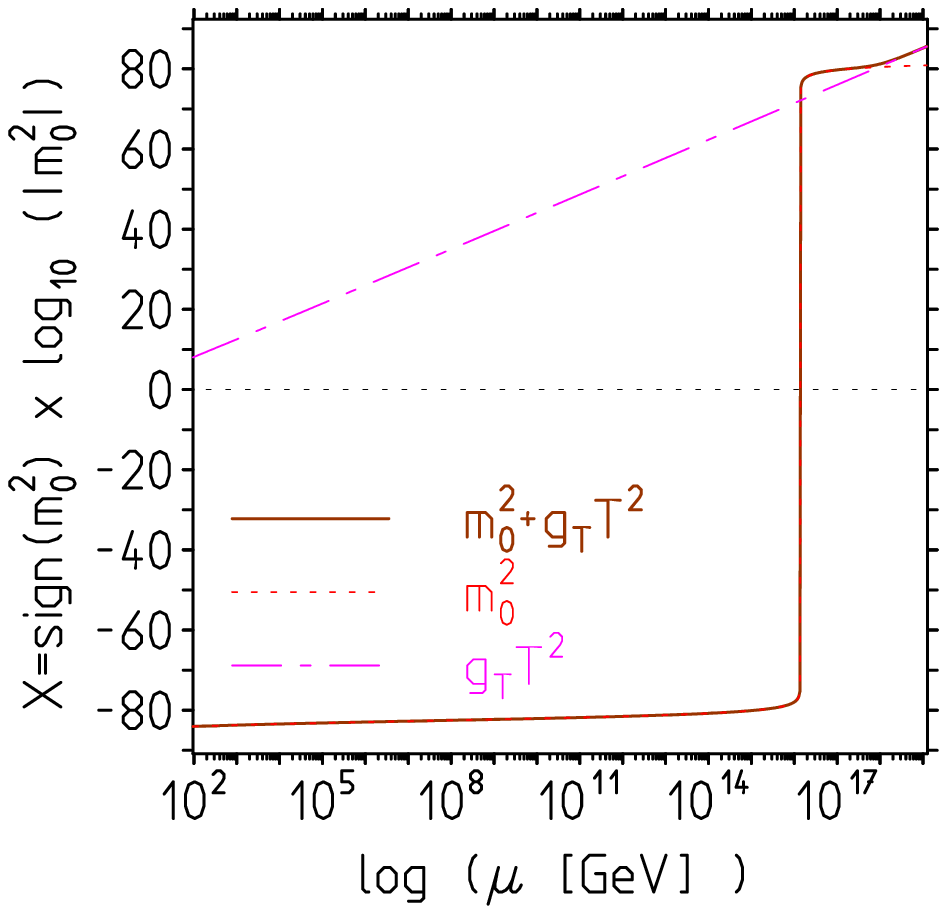}
\includegraphics[height=5.8cm]{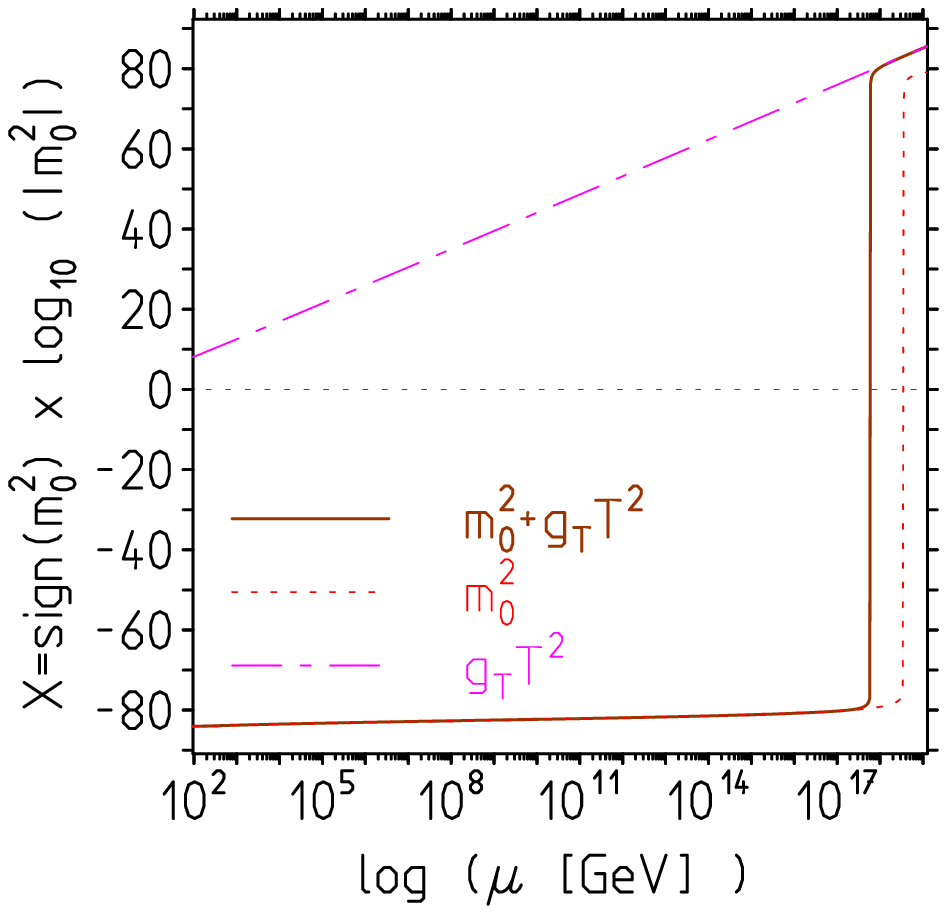}
\caption{The role of the Higgs in the finite temperature
SM. Left: for $\mu_0\sim 1.4 \power{16}~\gv$ ($M_H\sim
126~\gv$, $M_t\sim 173.5~\gv$).
Right: finite temperature delayed transition for
$\mu_0\sim 6 \power{17}~\gv$ ($M_H\sim 124~\gv$, $M_t\sim 175~\gv$),
the $m^2_0$ term  alone is flipping at about $\mu_0\sim 3.5 \power{18}~\gv$.}
\label{fig:jumpdmFT}
\end{figure}

In any case the zero of the coefficient function $C(\mu)$ triggers a phase
transition, which corresponds to a \textit{restoration of the
symmetry}. Indeed, there is a close relation between the Higgs
mechanism and the electroweak (EW) phase
transition~\cite{Dine:1992wr}. To this end we have to consider the
relevant finite temperature
effects~\cite{FiniteTemp1,FiniteTemp2,FiniteTemp3}, which are
dominating especially in the very early thermal evolution of the
universe at the hot big bang. Including the leading effect only, the
finite temperature effective potential reads
\bea
V(\phi,T) =
\frac{1}{2}\,\left(g_T\,T^2-\mu^2\right)\,\phi^2+\frac{\lambda}{24}
\,\phi^4 + \cdots\epo
\eea
The usual assumption is that the Higgs is in the broken phase
$\mu^2>0$ from the beginning at the big bang. The EW phase transition
is then taking place when the universe is cooling down below the
critical temperature $T_c=\sqrt{\mu^2/g_T}$, meaning
$g_T\,T^2-\mu^2<0$ when $T<T_c$. My analysis in contrast shows that
above the phase transition point $\mu_0$ the SM is in the symmetric phase
with $-\mu^2\to m^2_0=(m_H^2+\delta m_H^2)/2>0\,,$ and the EW phase
transition is essentially triggered by the Higgs mechanism, at least
it can happen only after the Higgs mechanism has taken place, thus
$\mu_{\rm EW}< \mu_{\rm HM}=\mu_0$. The relevant question here is
which of the terms $\delta m^2$ or $g_T\,T^2$ is leading in the
relevant epoch of in early universe? I find
$m^2_0(\mu=\MPl)\simeq 0.87\power{-3}\,\MPl^2$ such that
$T(\mu=\mu_0)\simeq 1.62\power{29}~\degK$ and $T(\mu=\MPl)\simeq
4.18\power{30}~\degK\epo$ We note that $T_{\mathrm{Pl}}\simeq
1.42\power{32}~\degK$ (temperature of the Big Bang). The coefficient $g_T$
is given by
$g_T=\frac{1}{4v^2}\,\left(2m_W^2+m_Z^2+2m_t^2+\frac12\,m_H^2\right)
=\frac{1}{16}\,\left[3\,g^2+{g'}^{2}+4\,y_t^2+\frac23\,\lambda
\right]\approx 0.0983\sim 0.1$ using the results of
Table~\ref{tab:params} at scale $\mpl$.
The dramatic jump in $m^2_0$ at $\mu_0$ in any case
drags the Higgs into the broken phase not far below $\mu_0$ as
illustrated in Fig.~\ref{fig:jumpdmFT}

\section{The Higgs hierarchy and its impact on inflation}
Cosmological 
inflation~\cite{Guth:1980zm,Starobinsky:1980te,Linde:1981mu,Albrecht:1982wi,Mukhanov:1981xt,Mukhanov:1985rz,Mukhanov:1990me}
requires an exponential growth of the Friedman-Robertson-Walker radius
of the universe $a(t)$, i.e.  $ a(t)\propto \E^{Ht}$ with
$H(t)=\dot{a}/a(t)$ the Hubble constant at cosmic time $t$. $\dot{X}$
denotes the time derivative of $X$. Inflation is able to solve the
\textit{flatness problem} (why is the actual energy density of the
universe so close to the critical density, the unique value which a
flat universe must have as a limiting case between the closed and the
open universes) and the \textit{horizon problem} (without inflation
what we seen when we look at the CMB
radiation, we would see a huge patch which at the time of last
scattering was outside the causal horizon, while the pattern is
observed to be uniform over all sky).  The
\textit{inflation term} comes in via the SM energy-momentum tensor
and adds to the r.h.s. of the Friedmann equation 
\bea
\ell^2\,\left(V(\phi)+\frac12 \,\dot{\phi}^2\right)\,,
\eea
where
$ \ell^2=8\pi G/3$, $ M_\mathrm{\rm Pl}=(G)^{-1/2}$ is the
Planck mass, $ G$ Newton's gravitational constant.

In the SM the Higgs contribution to the energy-momentum tensor
in terms of energy density and pressure amounts to
\bea
\rho_\phi=\frac12\,\dot{\phi}^2+V(\phi)\semis
p_\phi=\frac12\,\dot{\phi}^2-V(\phi)\epo
\eea
The second Friedman equation
$\ddot{a}/a=-\frac{\ell^2}{2}\,\left(\rho+3p\right)$ tells us that the
condition for growth $ \ddot{a}>0$ requires $p<-\rho/3$ and hence
$\frac12\dot{\phi}^2< V(\phi)$. CMB observations strongly favor the
slow-roll inflation $\frac12\dot{\phi}^2\ll
V(\phi)$ condition. Indeed the Planck mission
measured $w=p/\rho=-1.13^{+0.13}_{-0.10}$. The first Friedman equation reads
$\dot{a}^2/a^2+k/a^2=\ell^2\,\rho$ and may be written as
$H^2=\ell^2\,\left[V(\phi)+\frac12\,\dot{\phi}^2\right]=\ell^2\,\rho$,
while the field equation reads $\ddot{\phi}+3H\dot{\phi}=-V'(\phi)=-\D
V(\phi)/\D \phi\,$.
Note that the kinetic term $\dot{\phi}^2$ is controlled by
$\dot{H}=-\frac32 \ell^2\,\dot{\phi}^2=\ell^2\,\rho\,(q-1)$, i.e. by the
observationally controlled deceleration parameter $q(t)=-\ddot{a}a/\dot{a}^2$.

Inflation requires the presence of a dominating dark energy
contribution, characterized by the equation of state $p/\rho=-1$.
This is precisely what the SM in the symmetric phase
suggests. Provided the Higgs potential remains stable ($\lambda$
positive) a huge positive bare mass square at least naively supports
the Gaussian slow-roll inflation condition. Since both $\lambda$ and
$m^2$ for the first time are numerically fairly well known,
quantitative conclusions concerning the phenomenologically established
features of inflation should be possible solely on the basis of SM
properties. In a phase where the mass term is dominating, the behavior
is characterized by a free massive scalar field with potential
$V=\frac{m^2}{2}\,\phi^2$ such that
$H^2=(\dot{a}/a)^2=\frac{m^2}{6}\,\phi^2$ and
$\ddot{\phi}+3H\dot(\phi)=m^2\phi$ which is a harmonic oscillator with
friction. It tells us that the Higgs field is decaying more or less
rapidly, or looking back in time the Higgs field must grow
exponentially implying that $\phi$ must have been very large in the
early universe. This does not conflict with the expectation that the
SM Higgs field at low energies is of moderate size, as it is
renormalized by wave function renormalization factors which depend
logarithmically only on the renormalization scale and thus on the
cutoff. A huge Higgs field at early times in fact is crucial for SM
inflation to work, otherwise we would not get a sufficient amount
inflation. What also helps is the quartically enhanced cosmological
constant (CC) provided by the SM Higgs. In
Ref.~\cite{Jegerlehner:2014mua} we have shown that the corresponding
vacuum energy density is actually calculable by perturbative means,
with the result
\bea
V(0)=\frac{m^2}{2}\, \langle 0 | \phi^2 |0\rangle +
\frac{\lambda}{24}\,\langle 0 | \phi^4
|0\rangle=\frac{m^2}{2}\,\Xi+\frac{\lambda}{8}\,\Xi^2\,;\; \Xi=\frac{\mpl^2}{16\pi^2}\epo
\eea
With $m^2 \approx \delta m^2=\frac{\mpl^2}{32\pi^2}\,C(\mu)$ the
vacuum energy density reads
\bea
\rho_{\Lambda\,0}=\rho_{\Lambda\,} +\frac{\mpl^4}{(16\pi^2)^2}\,X(\mu) 
\label{finetuning}
\eea
with $X(\mu)=\frac18\,(2\,C(\mu)+\lambda(\mu))$. Thus $X(\mu)=0$ close
to the zero of $C(\mu)$, which takes relatively large negative values
at lower energies (see Fig.~\ref{fig:quaddivcoef})\footnote{The
non-vanishing $\langle 0 | \phi^2 |0\rangle$ also implies a shift of
the effective mass of the Higgs by ${m'}^2=m^2+\frac{\lambda}{2}\,\Xi\,.$}. Thus,
surprisingly, the cosmological constant and the Higgs mass term have
strongly correlated matching points, where the renormalized low energy
quantity coincides with the bare parameter and quadratic as well as
quartic cutoff effects are nullified not far from each other near the
EW phase transition point (see Ref.~\cite{Jegerlehner:2014mua} for
details).  Below the corresponding zeros the renormalized parameter
relations and parameter running applies and as low energy parameters
there is no reason why the renormalized quantities cannot be small. As
a result, the SM predicts a huge time-dependent CC, at $\mpl$
equivalent to $\rho_\phi\simeq V(\phi)\sim 2.77\,\mpl^4\sim
6.13\power{76}~\gv^4$, for the initial field value $\phi_i\simeq
4.51\,\mpl$ at Planck time $t_i=\tpl$, while the value observed today
is $\rho_{\rm vac}=\mu_\Lambda^4$ with $\mu_\Lambda\sim
0.002~\mbox{eV}$! The short distance versus long distance
``matching-patch'' separates the regime where we look at the bare
system form the ``illusory world'' we see at low energies where one
has lost the memory oft the cutoff.

The SM inflation pattern is impressively supported by observation,
most recently by the Planck 2013 results~\cite{PlanckResults}. The
cosmological constant is characterized by the equation of state
$w=p/\rho=-1$, and in our scenario is a prediction of the SM for times
before the phase transition when $\mu>\mu_0$. During the very early
inflation era, when $\frac12\dot{\phi}^2\ll V(\phi)$, the Higgs field
is decaying exponentially and a large Higgs field at the Planck scale
in not unnatural. In fact we need a huge field strength $\phi_i\simeq
4.51\,\mpl$ at Planck time $t_i=\tpl$, in order to get an amount of
inflation $N_e=\ln (a(t_e)/a(t_i))=\int_{t_i}^{t_e} H(t)\,\D t > 60$,
which is required as a minimum in order to solve the CMB horizon
problem. For the initial field value mentioned we obtain $N_e\approx
65$ and inflation ends at about $t_e\simeq 450\, \tpl$ with
$\phi_e\simeq 2\power{-3}\,\mpl$.  Inflation in any case would be
stopped by the \PT when $\mu=\mu_0$, however, due to the exponential
decay of the field inflation stops much earlier and field oscillations
set in before the phase transition is reached. In our scenario, in the
symmetric phase, the effective number of relativistic degrees of
freedom is $g_*(T)=g_B(T)+\frac78\,g_f(T)=102.75$ such that the Hubble
constant, during the very early radiation dominated era, reads $H=\ell
\sqrt{\rho}\simeq 1.66\,\left(k_B T\right)^2\sqrt{102.75}\,\mpl^{-1}$,
or at Planck time $H_i \simeq 16.83\,\mpl\approx 2.05\power{20}~\gv
\simeq 3.12\power{44}~\mbox{sec}^{-1}$ as an initial value, which
however decreases with $1/T^4$, such that the pure
inflation dark energy Hubble constant given by $H_\phi\simeq
\ell\sqrt{V(\phi)}\approx 4.81\,\mpl \approx 5.88\power{20}~\gv \simeq
8.93\power{43}~\mbox{sec}^{-1}$ becomes dominant and inflation sets
in. While $\phi$ is large the interaction term of the Higgs Lagrangian
will be dominating at first. As $\phi$ is decreasing the mass term
will be dominating for some time before inflation stops.

\section{Remarks concerning reheating and baryogenesis}
The four Higgses  near the Planck
scale have an effective mass about $m_{H0} \simeq 3.6\power{17}~\gv$ and thus can
be produces in processes like $WW\to HH$ or $t\bar{t} \to H$ at times
at and after the big bang. The big difference to standard big bang scenarios
is that the Higgses are primordial, i.e. they exist as modes in the
Planck medium in advance of being produced by high energy radiation
processes. A Higgs in this phase has a width dominated by
$H\to t\bar{t}$ decay, since direct couplings $HWW$ and $HZZ$ are absent
in the symmetric phase. One estimates
\bea
\Gamma_H\simeq \frac{m_{H0}}{16 \pi}\,N_c\,y^2_t(\mpl)
\simeq7.5\power{-3}\,m_{H0}\simeq 2.7\power{15}~\gv
\eea
yields a life time $\tau_H=1/\Gamma_H\simeq 2.5\power{-40}~\mbox{sec.}$
This is relatively large in terms of Planck times $\tpl \simeq 5.4
\power{-44}~\mbox{sec}$. It supports the possibility that the coupling
do not change immediately when dramatic cooling due to inflation takes
place. The SM predicts that
the Higgses produce top/anti-top quark radiation most abundantly. This
means that reheating is mainly provided by $H\to t\bar{t}$ decays. In
addition we estimate that $\Gamma_H
\ll H(t)=\dot{a}(t)/a(t)$  during inflation,
before the phase transition takes place. The energy density of
top/anti-top quarks produced by the Higgs decays satisfies the
conservation equation (see e.g. Ref.~\cite{Weinberg:2008zzc})
\bea
\dot{\rho}_t +3\,H\,(\rho_t+p_t) = \Gamma_H\,\rho_\phi\epo
\eea
Since the top quarks are relativistic $p_t=\rho_t/3$, and provided the energy density 
is still dominated by the inflaton, we can estimate
the maximum top radiation density. As a result one obtains
\bea
\rho_{t\,{\rm max}}&\leq&(3/8)^{8/5}\,t_i\,\Gamma_H\,\rho_\phi(t_i)
=0.139\,\left(\Gamma_H/H(t_i)\right)\,\rho_\phi(t_i) \nn \\
&\simeq&
0.139\,\frac{3\,\sqrt{3}\,y^2_t(\mpl)}{64\,\pi \sqrt{\pi}}\,\mpl^3\,m_{H0}
\simeq
1.6\power{71}~\gv^4,
\eea
with $t_i$ the Planck time.

Concerning the possibility of baryogenesis, baryon-number violating
interactions in the low energy effective SM (LEESM) scenario naturally are the close-by
dimension 6 effective four-fermion interactions discussed first by
Weinberg~\cite{Weinberg:1979sa}. Usually, it is assumed that some
unknown very heavy particle $X$ is responsible for baryogenesis. The
first stage is characterized by
\mbo{k_B T > m_X} when we have thermal equilibrium and $X$ production
and $X$ decay are in balance.  The second stage follows if \mbo{H
\approx \Gamma_X} and \mbo{k_B T < m_X} implying that $X$-production
is suppressed and the system moves out of equilibrium. Our $X$ is the
Higgs, with its known properties. Besides the predominant ``would-be
charged'' Higgs decays $H^+\to t\bar{b}$ and $H^-\to b\bar{t}$, with
rates proportional to $y_ty_b$, decays proportional to the
$CP$-violating CKM matrix-elements $V_{td}$ and $V_{ub}$ $H^+\to
t\bar{d},u\bar{b}$ and $H^-\to b\bar{u},d\bar{t}$ are important as a
condition for baryogenesis. At inflation times we have
\mbo{H^+\to t\bar{d}} with rate \mbo{\propto
y_ty_d\,V_{td}\sim 5.5\power{-8}\,(1-\rho-{\I\eta})} and
\mbo{H^-\to b\bar{u}} with rate \mbo{\propto y_by_u\,V_{ub}\sim
1.2\power{-9}\,(\rho-{\I\eta})}, where
\mbo{\rho=0.131,\,{\eta=0.345}}. 
The rates compare to the dominant $t$-mode\footnote{The
next-to-leading $b$-quark rates are reduced by the branching fraction
$4.4\power{-4}$ and the $\tau$-lepton rates are lower by
$2.2\power{-4}$.} with relative rate
\mbo{y_t^2\approx 0.123}. As mentioned before, matter
production is preferably into fermion pairs with the biggest Yukawa
couplings. After the EW phase transition
the now heavy states decay into the lighter ones, with the smaller
Yukawa couplings. Thus the major part of normal matter is produced via the
heavy states which are cascading down the CKM coupling
scheme. Apparently in such a scenario the system likely would
intermittently be far from equilibrium while approaching the EW phase
transition, and the dynamics behind could be important for the
explanation of the baryon-asymmetry. So, likely in this scenario the
origin of the baryon-asymmetry may have a different explanation than
thought so far.

\section{Conclusion}
The main conclusions have been given in the abstract already. Here we
would like to point out the importance of an extended analysis of the
possible consequences of the SM physics. One of our main assumptions
has been the one that physics beyond the SM is not needed to
understand the early universe. The point is that in the LEESM scenario
unseen physics can naturally be expected, however, it must be natural
in the sense of a low energy expansion. Grand unified theories as well
as a supersymmetrized SM are not natural, because they require an
improbably high amount of conspiracy of very many modes, while the
emergence of an extra $U(1)$ or a $SU(4)$ look much more natural. What
is also ruled out are additional fermion families. They definitely
would spoil the present interplay of couplings which make the
extrapolation up to the Planck scale working. 

We once more point out that there is no hierarchy problem in the
broken phase of the SM. All particle masses, the ones protected by
symmetries as well as the unprotected Higgs, are proportional to the
Higgs vacuum expectation value times a coupling which is subject to
logarithmic scale dependence only. The Higgs VEV is an order parameter
determined by collective long range properties of the system. If $v$
would be of order $\mpl$ the notion of spontaneous symmetry breaking
would be obsolete, since the symmetry would not be recovered at the
high scale. We should remember that the UV structure is the same in
the symmetric and in the broken phase, there cannot be any additional
UV cutoff sensitivity generated by the Higgs VEV. That $v$ is much
smaller than the cutoff, in principle can be checked by putting the SM
on a lattice of lattice spacing $a$ and then calculate $va$, which
should turn out to be extremely small $O(10^{-16})$, which is possible
if the temperature turns out to be very close just below the critical
temperature. This again is expected to be the result of the specific
conspiracy of the various couplings of the SM.

In any case, a super symmetric or any other extension of the SM cannot
be motivated by the (non-existing) hierarchy problem. The Higgs in a
supersymmetric extension of the SM cannot be the inflaton and provide
the necessary dark energy feeding inflation.

How do we get cold dark matter?  If the right-handed sterile singlet
neutrinos are Majorana particles exhibiting naturally a large Majorana
mass term not protected by any symmetry and not participating in the
Higgs spontaneous symmetry breaking could play a role here. Such
sterile Majorana neutrinos would naturally have masses of the order of
the Planck scale and therefore not affect the running parameters of
the SM, as they do not couple directly to any of the SM fields and as
they satisfy the decoupling theorem. At the same time they would provide
the seasaw mechanism which would explain the smallness of the neutrino
masses. But also an extra hidden $SU(4)$ could play a role here, by
forming stable bosonic quartet bound states. Cold dark matter could be
dominated by bound energy, similarly to the case of normal baryonic
matter with respect to QCD. Since $SU(4)$ bound states are bosonic
formation of structures and distribution of corresponding dark matter
would be very different form that of normal fermionic matter.

We also note that new physics like the existence of axions which could
play a key role in the issues of the strong CP problem, have a natural
place in a renormalizable low energy effective world.

As we have seen, a big issue is the very delicate conspiracy between SM
couplings. Therefore  precision determinations of parameters are more important
than ever and a real challenge for experiments at the LHC and at a future
ILC, which may improve substantially \mbo{\lambda}, \mbo{y_t} and
\mbo{\alpha_s}. But also low energy hadron facilities have to play an
important role as needed for a better control of the non-perturbative
hadronic effects in \mbo{\alpha(M_Z)} and \mbo{\alpha_2(M_Z)}. It is
important to note that, provided there is essentially no other stuff,
coming closer to the properties of the Planck ether is only possible
by pushing high precision physics. Thus higher order calculations and
high precision determinations of parameters are of paramount
importance. Note that the precise value of the top Yukawa contribution
plays a particular role for the precise location of the zero of the
coefficient of the quadratic divergence, as it is enhanced by a factor 6
relative to the Higgs self-coupling. Whether the Higgs is the inflaton
with the right properties depends crucially on the precise point in the
$(\lambda,y_t)$-plane. The window for this is very narrow.

Our analysis shows that the role of the Higgs is not just to provide
masses to SM particles, it also plays a key role in cosmology: for some
time at and after the big bang the Higgs is the only particle which
directly talks to gravity and directly takes part in the evolution of
the universe. It is the only SM particle which directly talks to the
vacuum in the early universe. Later, in the low energy phase,
contributions from the EW phase transition through the Higgs VEV and
from the QCD phase transition through quark and gluon condensates come
into play. This has to be investigated yet. Nevertheless, the Higgs
very likely is the object able to provide negative pressure and likely
is responsible for blowing continuously energy into the expanding
universe according to the established dark energy which still is
existing today. While the Higgs likely was playing a dominating role in
shaping the early universe, in our present world the Higgs hides itself so
much that it took decades to actually find it, after theorists had
proposed it as being the  source of the masses of the SM particles.

\noindent
{\bf Acknowledgments} \\ Thanks to the organizers for the
invitation and support to the XX Epiphany Conference, 
and for giving me the opportunity to present
this talk.

\end{document}